\documentclass[10pt,twocolumn,twoside]{IEEEtran}
\usepackage{amsmath,graphicx}
\usepackage{amssymb}
\usepackage{amsfonts}
\usepackage{mathtools}
\usepackage{multirow}
\usepackage{bm}
\usepackage{nicefrac}
\usepackage{tabularx}
\usepackage{booktabs}
\usepackage{xcolor}
\usepackage{pifont}
\usepackage{amsthm}
\usepackage{extarrows}
\usepackage{hyperref}
\usepackage{balance}
\usepackage{subcaption}
\usepackage{caption}
\usepackage[ruled,vlined]{algorithm2e}
\usepackage{cite}
\usepackage{graphicx}
\usepackage{verbatim}
\usepackage{url}
\usepackage{stfloats}
\usepackage{textcomp}
\usepackage{array}
\usepackage{tabulary}
\usepackage{color, colortbl}
\usepackage{hhline}

\newtheorem{theorem}{Theorem}[section]

\SetKwInput{KwInput}{Input}                
\SetKwInput{KwOutput}{Output}              

\def\x{{\mathbf x}}
\def\w{{\bm{\theta}}}

\def\s{{\mathbf s}}
\def\lp{\left(}
\def\rp{\right)}

\def\L{{\cal L}}
\def\D{{\cal D}}
\def\R{{\mathbb R}}
\def\E{{\mathbb E}}

\newcommand{\T}{\mathrm{T}}

\def\R{\mathbb{R}}

\def\es{\widehat{\mathbf{s}}}

\def\n{\mathbf{n}}

\def\en{\widehat{\mathbf{n}}}

\def\s{\mathbf{s}}

\def\mix{\mathbf{m}}

\def\tmix{\widetilde{\mathbf{m}}}

\def\tes1{\widetilde{\mathbf{s}}_{1}}
\def\tes2{\widetilde{\mathbf{s}}_{2}}

\def\tes{\widetilde{\mathbf{s}}}

\def\ten1{\widetilde{\mathbf{n}}_{1}}
\def\ten2{\widetilde{\mathbf{n}}_{2}}

\def\ten{\widetilde{\mathbf{n}}}

\makeatletter
\def\hlinewd#1{%
  \noalign{\ifnum0=`}\fi\hrule \@height #1 \futurelet
   \reserved@a\@xhline}
\makeatother

\begin{document}
\title{RemixIT: Continual self-training of speech enhancement  models via bootstrapped remixing}

\author{
Efthymios Tzinis \IEEEmembership{Graduate Student Member, IEEE},
Yossi Adi \IEEEmembership{Member, IEEE},
Vamsi K. Ithapu \IEEEmembership{Member, IEEE},\\ 
Buye Xu \IEEEmembership{Member, IEEE},
Paris Smaragdis \IEEEmembership{Fellow, IEEE},
Anurag Kumar \IEEEmembership{Member, IEEE} \\
\thanks{E. Tzinis is with the University of Illinois at Urbana-Champaign, Urbana, Illinois, USA (email: etzinis2@illinois.edu). Part of the work was done while E. Tzinis was an intern at Meta Reality Labs Research. Y. Adi is with Meta AI Research, Tel Aviv, Israel (email: adiyoss@fb.com). V. Ithapu and B.Xu are with Meta Reality Labs Research, Redmond, WA, USA (email: ithapu, xub@fb.com). P. Smaragdis is with the University of Illinois at Urbana-Champaign, Urbana, Illinois, USA (email: paris@illinois.edu). A. Kumar is with Meta Reality Labs Research, Redmond, WA, USA (email: anuragkr@fb.com).}
}

\markboth{Journal of \LaTeX\ Class Files,~Vol.~14, No.~8, August~2021}%
{Shell \MakeLowercase{\textit{et al.}}: A Sample Article Using IEEEtran.cls for IEEE Journals}

\IEEEpubid{0000--0000/00\$00.00~\copyright~2022 IEEE}


\maketitle

\begin{abstract}
We present \textit{RemixIT}, a simple yet effective self-supervised method for training speech enhancement without the need of a single isolated in-domain speech nor a noise waveform. Our approach overcomes limitations of previous methods which make them dependent on clean in-domain target signals and thus, sensitive to any domain mismatch between train and test samples. \textit{RemixIT} is based on a continuous self-training scheme in which a pre-trained teacher model on out-of-domain data infers estimated pseudo-target signals for in-domain mixtures. Then, by permuting the estimated clean and noise signals and \textit{remixing} them together, we generate a new set of bootstrapped mixtures and corresponding pseudo-targets which are used to train the student network. Vice-versa, the teacher periodically refines its estimates using the updated parameters of the latest student models. Experimental results on multiple speech enhancement datasets and tasks not only show the superiority of our method over prior approaches but also showcase that \textit{RemixIT} can be combined with any separation model as well as be applied towards any semi-supervised and unsupervised domain adaptation task. Our analysis, paired with empirical evidence, sheds light on the inside functioning of our self-training scheme wherein the student model keeps obtaining better performance while observing severely degraded pseudo-targets.
\end{abstract}

\begin{IEEEkeywords}
Self-supervised learning, speech enhancement, semi-supervised self-training, zero-shot domain adaptation.
\end{IEEEkeywords}

\section{Introduction}
\label{sec:intro}

One of the most fundamental problems in audio processing is speech enhancement, where the goal is to isolate and reconstruct the clean speech component from a noisy input recording \cite{benesty2006_general_speech_enhancement}. Several studies have shown that employing such denoising models as front-ends could be useful for building robust automatic speech recognition (ASR) \cite{weninger2015SE_w_LSTM_robustASR,sainath2017multichannel_SE_for_ASR} and speaker recognition \cite{taherian2020robustSE} systems. The universal applicability of neural networks has proven to be beneficial for a variety of signal processing problems, including speech enhancement. Sophisticated architectures such as convolutional networks \cite{choi2018phaseaware_speechenhancement,zhao2018twostagefor_speechenhancement,defossez20_interspeech,hao2021fullsubnet}, recurrent processing \cite{zhao2018early_RCNN_speechenhancement} self-attention \cite{koizumi2020speech,Isik2020PoCONetSpeechEnhacement,pandey2021dense}, generative adversarial networks \cite{pascual2017segan,fu2019metricgan} as well as variational auto-encoders \cite{bando2018statistical_speeechenhancemnt_VAE_NMF}, to name a few. Despite the effectiveness of the aforementioned approaches in cases where large amounts of in-domain training paired data are available, real-world applications necessitate the need for developing robust algorithms to train these models with in-the-wild mixtures.

In the context of speech enhancement, self-supervised learning (SSL) or unsupervised methods differ from semi-supervised ones \cite{xia2021incorporating} in the sense that the former do not have access to clean target signals. Orthogonal to these concepts, self-training refers to algorithms which are able to train a new model (student) based on pseudo-targets provided by a previously fitted model (teacher). Under this unified terminology, the proposed \textit{RemixIT} framework can also be viewed as an unsupervised self-training algorithm when only unsupervised data are used to pre-train the teacher model.

Recent studies have shown that speech representations could be self-learned and be used later for other downstream audio processing tasks \cite{wav2vec_2.0,wang2020unsup_pretrain_speech_enc_masked_recosntruction,liu2021tera_SSL_transformer_encoder_speech}. However, in real-world settings, the speech recordings are degraded with additive noise, thus, self-learning robust embeddings becomes particularly challenging and demands the adaptation to the input noise distribution \cite{liao2018noise_adaptive_speech_enhancement_domain_adv_training}. Several unsupervised speech denoising algorithms have been proposed by identifying and training with relatively clean segments of the noisy speech mixture \cite{zezario2020self_linearfilter_highsnr, sivaraman2021personalizedSE_SSL_dataPurification}, using ASR losses \cite{subramanian2019speech_enhancement_using_end2endASR,trinh2021unsupervised_speechenhancement_ASR_disentanglement_losses} exploiting visual cues \cite{cheng2021multimodal_SSL_curriculumlearn_SE}, and harnessing the spatial separability of the sources using mic-arrays \cite{li2019multichannelSE_unsupbeamforming, shimada2019unsupervisedSE_multich_NMFinfrmd_beamforming}. Mixture invariant training (MixIT) \cite{mixit} enables unsupervised training of separation models only with real-world single-channel recordings by generating artificial mixtures of mixtures and estimating the independent sources. Although MixIT has been proven successful for various speech enhancement tasks \cite{mixit,fujimura2021noisyTargetTRaining,tzinis2021separate}, MixIT assumes access to in-domain noise samples which restricts its universal applicability. Overcoming the latter constraint by injecting additional out-of-domain (OOD) noise sources to the input mixture of mixtures \cite{saito2021trainingSEsystemsWNoisyDatasets} further alters the input signal-to-noise ratio (SNR) distribution and its performance depends heavily on the distribution shift between the injected and real noise distributions. Thus, developing a SSL algorithm which does not depend on external modality information nor assumptions about in-domain data remains a challenging problem. 

\IEEEpubidadjcol

On the other hand, several self-training strategies have emerged and showed promising results in classification tasks using convex combinations of labeled and unlabeled data (e.g. Mixup \cite{zhang2018mixup}) but have also been successfully applied to several audio tasks \cite{aihara2019teacherStudentDeepClustering,hsu2021hubert}. In \cite{zhang2021teacher}, a student model with a smaller number of estimated sources has been trained on a subset of outputs of a pre-trained MixIT model to solve the input SNR distribution mismatch. Furthermore, a student model could also perform test-time adaptation by using the teacher's estimated waveforms as targets \cite{zeroshot_testTimeAdaptationSE}. However, those approaches enforce only the consistency of the student's predictions over a frozen teacher's output pseudo-targets whereas other studies have shown that one can obtain significant gains using unsupervised data augmentation \cite{xie2020unsupervised}, averages of losses over multiple predictions \cite{remixmatch_distribution_match_anchoring}, or their combination \cite{fixmatch}. 


The student-teacher framework for singing-voice separation in \cite{wang2021semiSupSingVoiceSepNoisySelfTraining} bears the closest similarity to our work. The proposed setup assumes teacher pre-training on supervised OOD data, performing inference on the in-domain noisy dataset and storing the new pseudo-labeled dataset. At a second step, a student network is trained on randomly mixed estimated sources that score above a pre-defined confidence quality threshold. Unfortunately, if the teacher's estimates have low SNR and/or the threshold is not picked wisely then the student model would also perform poorly. In contrast, some of the most successful self-training approaches propose to iteratively update the teacher's weights using an exponential moving average scheme \cite{mean_teachers_better_role_models, higuchi2021momentumPseudoLabelingforASR, xu2020iterative} or sequentially update the teacher with the weights from a more expressive noisy student \cite{xie2020self}.

\begin{figure*}[t!]
    \centering
      \includegraphics[trim={4cm 0 2.5cm 0},width=0.9\linewidth]{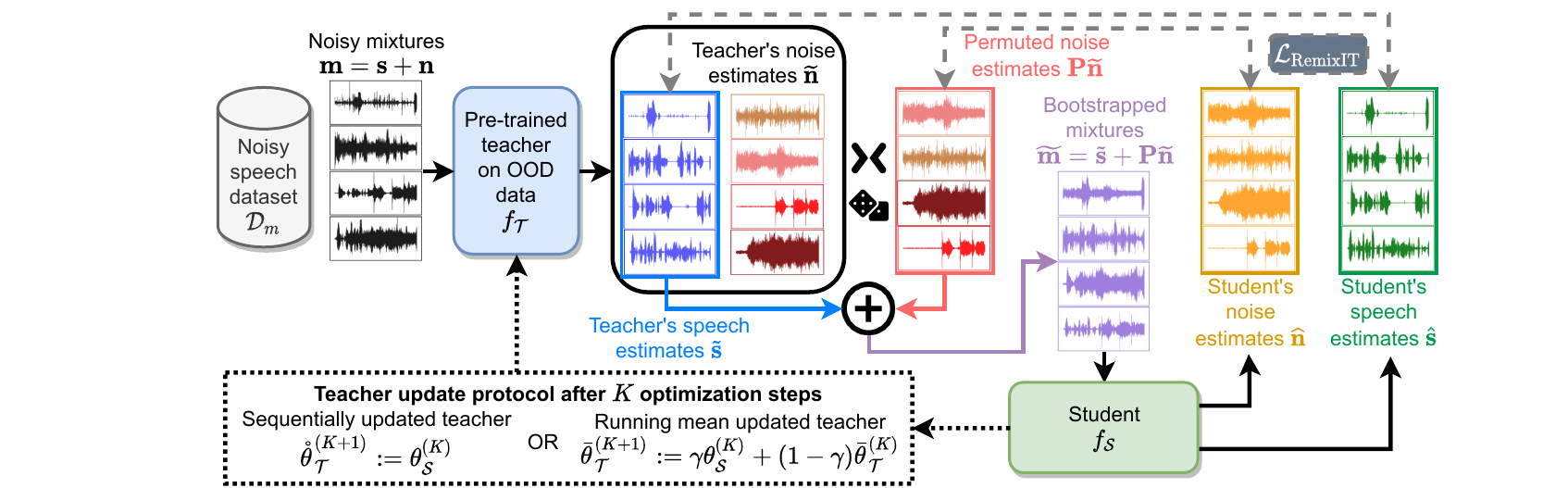}
      \caption{\textit{RemixIT} self-training procedure with a batch size of 4 noisy mixtures. A teacher speech enhancement model $f_{\mathcal{T}}$ is pre-trained in a supervised or unsupervised way on out-of-domain (OOD) data and performs inference on a batch of noisy mixtures sampled from the in-domain noisy speech dataset $\mix \sim \D_m$. The randomly permuted teacher's noise estimates $\textbf{P} \ten$ are added together with the teacher's speech estimates $\tes$ to form the bootstrapped mixtures $\tmix$ which are fed to the student speech enhancement network $f_{\mathcal{S}}$. The student is trained by regressing over the teacher's estimated sources which are now used as pseudo-targets under a specified signal-level loss function. After repeating the overall process for $K$ optimization steps, the teacher model may be updated using the student's weights in a continuous self-training scheme.}
      \label{fig:overall_remixit}
    \vspace{-10pt}
\end{figure*}

In this work, we propose \textit{RemixIT} which is based on several aforementioned state-of-the-art SSL strategies for pseudo-labeling and continual training while also providing a novel technique for training speech enhancement models with OOD data. Our method trains a student model using self-augmented mixtures generated by permuting and remixing the teacher's estimates and using them as pseudo-targets for regular regression. Moreover, \textit{RemixIT} treats self-training as a lifelong process while continually updating the teacher model using the student's weights that consequently leads to faster 
and more robust convergence. \textit{RemixIT} is the first method that:
\begin{itemize}
    \item Performs self-supervised learning using only in-domain mixture datasets and OOD noise sources (e.g. MixIT pre-trained teacher with an OOD dataset).
    \item Yields state-of-the-art results on several unsupervised and semi-supervised denoising tasks without the need of clean speech waveforms or ad-hoc filtering procedures.
    \item Has strong theoretical and empirical evidence of why it works under various noise levels.
    \item Is able to leverage huge amounts of unsupervised data and generalize in diverse training and adaptation scenarios.
\end{itemize}

\section{RemixIT Method}
\label{sec:remix_method}
\textit{RemixIT} trains a speech enhancement model to isolate the clean speech signal from its noisy observation. In general, we train a separation model $f$ which outputs $M$ source waveforms for each input noisy speech recording with $T$ time-domain samples. Thus, given as input a batch of $B$ input waveforms $\x \in \R^{B \times T}$ the network estimates all sound sources:
\begin{equation}
\label{eq:basic_estimation}
    \begin{gathered}
    \es, \en = f(\x; \w),\enskip
    \x = \s + {\textstyle \sum}_{i=1}^{M-1} [\n]_i = \es + {\textstyle \sum}_{i=1}^{M-1} [\en]_i , 
    \end{gathered}
\end{equation}
where $\es, \, \s \in \R^{B \times T}$, $\en, \, \n \in  \R^{(M-1) \times B \times T}$, $\w$ are: the estimated speech signal, the clean speech target, the estimated noise signal, the noise target and the parameters of the model, respectively. We force the estimated sources $\es$ and $\en$ to add up to the initial input mixtures $\x$ by using a mixture consistency projection layer \cite{wisdom2019differentiableMixtureConsistency}. We portray the inference and self-training aspects of \textit{RemixIT} in Figure \ref{fig:overall_remixit}, summarize it in Algorithm \ref{alg:remixit} and analyze it in depth in Section \ref{sec:remix_method:remixit}. For completion, we highlight how \textit{RemixIT} differs from fully supervised training (assumes access to clean in-domain speech) and previous state-of-the-art semi-supervised training methods (MixIT assumes access to isolated in-domain noise recordings) in Sections \ref{sec:remix_method:supervised_training} and \ref{sec:remix_method:mixit}, respectively. 

\begin{algorithm}[htb]
\SetAlgoLined
    $\w^{(0)}_{\mathcal{T}} \gets \textsc{Pretrain\_Teacher}(f_{\mathcal{T}}, \D')$ \\
    $\w_{\mathcal{S}} \gets \textsc{Initialize\_Student}(f_{\mathcal{S}})$ \\
 \For{$k = 0$; $k{+}{+}$; while $k <= K$}{
 \For{\textsc{Sample\_Batch} $\mix \in \D_m, \enskip \mix \in \R^{B \times T}$}{ 
    $ \tes, \ten \gets f_{\mathcal{T}}(\mix; \w^{(k)}_{\mathcal{T}}) $ \tcp{\footnotesize Teacher's estimates}
    $\tmix = \tes + \textbf{P} \ten$ \tcp{\footnotesize Bootstrapped remixing}
    $\es, \en  \gets f_{\mathcal{S}}(\tmix; \w^{(k)}_{\mathcal{S}}) $ \tcp{\footnotesize Student's estimates}
    $\mathcal{L}_{\operatorname{RemixIT}} = \sum_{b=1}^B \left[ \L(\widehat{s}_b, \widetilde{s}_b) + \L(\widehat{n}_b, \left[\textbf{P} \ten \right]_b) \right]$ \\
    $\w_{\mathcal{S}} \gets \textsc{Update\_Student}(\w_{\mathcal{S}}, \nabla_{\w_{\mathcal{S}}} \mathcal{L}_{\operatorname{RemixIT}})$
 } 
 $\w^{(k+1)}_{\mathcal{T}} \gets \textsc{Update\_Teacher}(\w^{(k)}_{\mathcal{T}}, \w_{\mathcal{S}})$
}
 \caption{\textsc{RemixIT} for the noisy dataset $\D_m$.}
 \label{alg:remixit}
\end{algorithm}  
\subsection{Supervised training}
\label{sec:remix_method:supervised_training}
Supervised training is the straightforward way of training speech enhancement models. It assumes access to both in-domain clean speech recordings, $\s \sim \D_s$, as well as noise sources drawn from $\n \sim \D_n$. Synthetic mixtures are generated at each training step $\mix = \s + \n$, by sampling a batch of clean speech recordings $\s \sim \D_s$ and a batch of isolated noise samples $\n \sim \D_n$, which are then fed to the separation model $f$. For a sampled batch of $B$ input mixtures, the model predicts $M=2$ sources for each input mixture ($\es, \, \en = f(\x; \w)$) and the following targeted loss function is minimized:
\begin{equation}
\label{eq:supervised_loss}
    \begin{aligned}
    \mathcal{L}_{\operatorname{Supervised}} = \textstyle \sum_{b=1}^B \left[ \L(\widehat{s}_b, s_b) + \L(\widehat{n}_b, n_b) \right],
    \end{aligned}
\end{equation}
where $\L$ is any desired signal-level loss function used to penalize the reconstruction error between the estimates and their corresponding targets. However, this training process is completely dependent on the availability of clean speech and noise sources to capture the real-world mixture distribution, making the model vulnerable to a performance decline under unseen test conditions. This necessitates the development of SSL and adaptation techniques for speech enhancement.  

\subsection{Mixture invariant training (MixIT)}
\label{sec:remix_method:mixit}
MixIT \cite{mixit} is a simple yet effective idea for training a separation model using artificial mixtures of mixtures (MoMs). In essence, MixIT assumes availability of two sources of data during training, $\D_m$ which consists of mixtures of speech and a noise source and $\D_n'$ which contains noise recordings from a single noise source. The training process boils down to sampling a batch of noisy speech recordings $\mix \sim \D_m$ (where $\mix=\s + \n^{(1)}$), and mixing them with another batch of isolated noise recordings $\n^{(2)} \sim \D_n'$. Note that the true noise distribution of the real-world $\D_n^*$ ($\n^{(1)} \sim \D_n^*$) is unknown and not necessarily same as the one available $\D_n'$. 
The separation model $f_{\mathcal{M}}$ is trained using the synthetic batch of input-MoMs $\x = \s + \mathbf{n}_1 + \mathbf{n}_2$ and tries to reconstruct $M=3$ sources $\es,\, \en^{(1)}, \, \en^{(2)} = f_{\mathcal{M}}(\x; \w_{\mathcal{M}})$, by minimizing the following permutation invariant \cite{Yu2017PIT} loss function:
\begin{equation}
\label{eq:mixit_loss}
    \begin{aligned}
    \mathcal{L}_{\operatorname{MixIT}}^{(b)} = 
    \underset{\bm{\pi} \in \mathcal{P} }{\min} 
    \left[ \L(\widehat{s}_b + \widehat{n}_{b}^{(\pi_1)}, m_b) + \L(\widehat{n}_{b}^{(\pi_2)}, n_{b})  \right], 
    \end{aligned}
\end{equation}
where $b$ is the batch's index and $\mathcal{P} := \{(1,2), (2,1)\}$ is the set of permutations between the model's noise output slots. One could also use a probabilistic assignment of the noise estimates $\widehat{n}_{b}^{(\pi_1)}$, $\widehat{n}_{b}^{(\pi_2)}$ to avoid emerging problems with the complex permutation invariant landscapes \cite{Yousefi2019}.

If the noise sources are independent from each other and the clean speech component, then the model can learn to minimize this loss by reconstructing the mixture using its first estimated slot and either one of the two noise slots available. Although MixIT has been proven effective for various simulated speech enhancement setups \cite{fujimura2021noisyTargetTRaining,tzinis2021separate}, the assumption about having access to a diverse set of in-domain noise recordings from $\D_n'$ which aptly captures the true distribution of the present background noises $\D_n^*$ make it impractical for many real-world settings. To this end, other works \cite{saito2021trainingSEsystemsWNoisyDatasets,trinh2021unsupervised_speechenhancement_ASR_disentanglement_losses} have tried to deal with the distribution shift between the on-hand noise dataset $\D_n$ and the actual noise distribution $\D_n^*$ in order to avoid the need of in-domain noise samples. Specifically, \cite{saito2021trainingSEsystemsWNoisyDatasets} proposes to use extra noise injection from an OOD distribution and in \cite{trinh2021unsupervised_speechenhancement_ASR_disentanglement_losses} ASR and disentanglement losses have been proposed. However, the performance of the former method still depends heavily the level of distribution shift between the actual noise distribution $\D_n^*$ and $\D_n$ while the latter method is more restrictive since it requires large pre-trained ASR models.

\subsection{RemixIT: Self-training with bootstrapped remixing}
\label{sec:remix_method:remixit}
In contrast to the aforementioned two training procedures which require in-domain ground truth signals (e.g. supervised training requires clean speech samples from $\s \sim \D_s$ as well as access to in-domain noise recordings sources drawn from $\n \sim \D_n$ while MixIT requires only isolated in-domain noise waveforms), \textit{RemixIT} does not depend on any other in-domain information besides the mixture dataset $\D_{m}$. Specifically, our method utilizes a student-teacher framework where the teacher's noise estimates are randomly permuted in a mini-batch sense and remixed with the teacher's speech estimates to create bootstrapped mixtures. A student model is trained using as input the bootstrapped mixtures and regressing over the teacher's pseudo-target signals using a regular supervised loss at every optimization step (for a succinct description of the training procedure please see Algorithm \ref{alg:remixit}). \textit{RemixIT} also enjoys a continual refinement of the noisy pseudo-target signals, after a few optimization steps, where the student model weights are used to update the teacher network as it is illustrated in Figure \ref{fig:overall_remixit}.

\subsubsection{RemixIT's teacher-student framework}
\label{sec:remix_method:remixit:student_teacher}
For the initial teacher model, \textit{RemixIT} can use any speech enhancement model pre-trained on an OOD dataset $\D'$ which outputs the speech component and one or more noise estimated waveforms (see specification in Equation \ref{eq:basic_estimation}). To this end, \textit{RemixIT} materializes into semi-supervised domain adaptation if the teacher was trained using a supervised loss and into a SSL training scheme if the teacher was trained using MixIT. 

Formally, given an batch of in-domain noisy mixtures $\mix = \s + \n \in \R^{B \times T}, \enskip \mix \sim \D_m$, the teacher model estimates the speech and the noise components as follows: 
\begin{equation}
\label{eq:teacher_inference}
    \begin{gathered}
    \tes, \, \ten  = f_{\mathcal{T}}(\x; \w^{(k)}_{\mathcal{T}}), \enskip
    \mix  = 
    \tes + {\textstyle \sum}_{i=1}^{M-1} [\textbf{n}']_i \\ 
    \end{gathered}
\end{equation}
where $\w^{(k)}_{\mathcal{T}}$ denotes the parameters of the teacher model at the $k$-th optimization step. The second equation holds because we enforce mixture consistency. A MixIT pre-trained model would estimate $M=3$ sources and we can easily get a consolidated noise estimate by summing the two latter noise estimated waveforms, namely, $\ten = \sum_{i=1}^{M-1} [\textbf{n}']_i$. Notice that the teacher model $f_{\mathcal{T}}$ does not need to be identical through the whole training process and could be updated using any user-specified protocol which results in a separation model that respects the constraints defined in Equation \ref{eq:teacher_inference}. The teacher's estimates within a batch of size $B$ are used to generate the bootstrapped mixtures $\tmix$ by remixing the estimated speech and noise sources in a random order:
\begin{equation}
\label{eq:bootstrapped_mixtures}
    \begin{gathered}
    \tmix = \tes + \ten^{(\textbf{P})} \in \R^{B \times T}, \enskip \ten^{(\textbf{P})} = \textbf{P} \ten, \enskip \textbf{P} \sim \bm{\Pi}_{B \times B},
    \end{gathered}
\end{equation} 
where $\textbf{P}$ is drawn uniformly from the set of all $B \times B$ permutation matrices and is used to produce the permuted noise sources $\ten^{(\textbf{P})}$. The original teacher's speech estimates $\tes$ and the permuted noise sources $\ten^{(\textbf{P})}$ are now used as target pairs to train the student model $f_{\mathcal{S}}$ on the newly generated batch of bootstrapped mixtures $\tmix$ as shown below:
\begin{equation}
\label{eq:student_train}
    \begin{gathered}
    \es, \, \en = f_{\mathcal{S}}(\tmix; \w^{(k)}_{\mathcal{S}}), \enskip \es, \, \en \in \R^{B \times T} \\
    \mathcal{L}_{\operatorname{RemixIT}}^{(b)} = \L(\es_b, \tes_b) + \L(\en_b,  \textbf{P} \ten_b), \enskip  b \in \{1, \dots, B\}. 
    \\
    \mathcal{L}_{\operatorname{RemixIT}} = \textstyle \sum_{b=1}^B \left[ \L(\es_b, \tes_b) + \L(\en_b, \textbf{P} \ten_b) \right] \\
    \end{gathered}
\end{equation}
The loss function used is similar to a supervised setup (see Equation \ref{eq:supervised_loss}) but instead of ground-truth clean source waveforms, we use the noisy estimates, $\tes$ and $\ten^{(\textbf{P})}$, provided by the teacher network. If the signal-level loss function $\L$ also minimizes the Euclidean norm between the estimated signals and the target signals, the proposed cost function $\mathcal{L}_{\operatorname{RemixIT}}$ enjoys several convergence properties which enable our method to learn in a robust SSL fashion even in cases where the teacher's estimates are not close to the ground-truth source waveforms (see Section \ref{sec:remix_method:remixit:error_analysis}).

Lastly, \textit{RemixIT} refines the estimates of the teacher network $f_{\mathcal{T}}$ using the weights from the latest available student models. The continual update protocols used in this study are the sequential and the running moving average update protocols which are explained in detail in Section \ref{sec:exp_fram:remixit_coinfigs}.

\subsubsection{Error analysis under the Euclidean norm}
\label{sec:remix_method:remixit:error_analysis}
In each optimization step, \textit{RemixIT} tries to minimize a signal-level loss function between the student's estimates and the teacher's pseudo-targets. Since we are mostly interested in denoising, we focus on the speech estimates of the teacher and the student networks with initial mixtures $\mathbf{M}$ and the bootstrapped mixtures $\widetilde{\mathbf{M}}$ as inputs, respectively. These estimates can also be expressed in the following way as random variables:
\begin{equation}
\label{eq:student_teacher_estimates_RVs}
    \begin{gathered}
    \widetilde{\mathbf{S}} = f^{(\widetilde{s})}_{\mathcal{T}} (\mathbf{M} = \mathbf{S} + \mathbf{N}; \,\w^{(k)}_{\mathcal{T}} ) ,  \enskip \mathbf{M}  \sim \D_m  \\
    \widehat{\mathbf{S}} = f^{(\widehat{s})}_{\mathcal{S}} \, (\widetilde{\mathbf{M}} = \widetilde{\mathbf{S}} + \widetilde{\mathbf{N}}^{(\textbf{P})}; \w^{(k)}_{\mathcal{S}} ).
    \end{gathered}
\end{equation}
Now, the teacher's $\widetilde{\mathbf{R}}_{\cal T}$ and student's $\widehat{\mathbf{R}}_{\cal S}$ errors w.r.t. the initial clean targets $\mathbf{S}$ are the following conditional probabilities: 
\begin{equation}
\label{eq:student_teacher_errors}
    \begin{gathered}
    \widetilde{\mathbf{R}}_{\cal T} = \widetilde{\mathbf{S}} - \mathbf{S},  \enskip \enskip \widetilde{\mathbf{R}}_{\cal S} = \widehat{\mathbf{S}} - \mathbf{S} \\ 
    \widetilde{\mathbf{R}}_{\cal T} \sim P(\widetilde{\mathbf{R}}_{\cal T} | \mathbf{S}, \mathbf{N}) , \enskip
        \widehat{\mathbf{R}}_{\cal S} \sim P(\widehat{\mathbf{R}}_{\cal S} | \widetilde{\mathbf{S}}, \widetilde{\mathbf{N}}, \textbf{P}).
    \end{gathered}
\end{equation}
Using a signal-level loss $\L$ that minimizes the squared error between the estimated and the target signals in Equation \ref{eq:student_train} and assuming unit-norm estimated and target signals $||s||=||\widetilde{s}||=||\widehat{s}||=1$, \textit{RemixIT} loss function becomes equivalent to minimizing the following expression:
\begin{equation}
\label{eq:equiv_student_train}
    \begin{gathered}
    \mathcal{L}_{\operatorname{RemixIT}} \propto 
    \E [ || \widehat{\mathbf{S}} - \widetilde{\mathbf{S}} ||^2_2 ] =  
    \E [ || ( \widehat{\mathbf{S}} - \mathbf{S} ) - ( \widetilde{\mathbf{S}} - \mathbf{S} ) ||^2_2 ] \\
    =
    \underbrace{\E \left[ || \widehat{\mathbf{R}}_{\cal S} ||^2_2 \right]}_{\text{Supervised Loss}}
    + \underbrace{\E \left[ || \widetilde{\mathbf{R}}_{\cal T} ||^2_2 \right]}_{\text{Constant w.r.t. }\w_{\cal S}}
    - 2 \underbrace{\E \left[ \langle \widehat{\mathbf{R}}_{\cal S}, \widetilde{\mathbf{R}}_{\cal T} \rangle \right]}_{\text{Errors' correlation}}
    \end{gathered}
\end{equation}
Ideally, this loss could lead to the same optimization objective with a supervised setup if the last inner-product term was zero since the middle term becomes zero when computing the gradient w.r.t. the student's parameters $\w_{\cal S}$. $\langle \widehat{\mathbf{R}}_{\cal S}, \widetilde{\mathbf{R}}_{\cal T} \rangle = 0$ could be achieved if the teacher produced outputs indistinguishable from the clean target signals or the conditional error distributions in Equation \ref{eq:student_teacher_errors} were independent. Intuitively, as we continually update the teacher model and refine its estimates, we minimize the norm of the teacher error which leads to higher fidelity reconstruction from the student (for further analysis of how the student learns to perform better than its teacher and for experimental validation of this claim we refer the reader to Section \ref{sec:results:student_learning_progression}).

Additionally, the bootstrapped remixing process forces the errors to be more uncorrelated since the student tries to reconstruct the same clean speech signals $\s$, similar to its teacher, but under a different mixture distribution. Formally, the student tries to reconstruct $\s$ when observing the bootstrapped mixtures $\widetilde{\mix} = \tes + \widetilde{\n}^{(\textbf{P})}$ while the teacher tries to reconstruct $\s$ only from the initial input mixtures $\mix=\s+\n$. This phenomenon becomes apparent if we focus on the reconstruction of a single speech signal $s^*$ from the teacher and the student networks. In essence, we use the teacher network to provide an estimated $\widetilde{s}^*$ from the corresponding mixture $m^*$ and some perturbed noise sources $\widetilde{n}_b', \enskip \forall b \in \{1, \dots, B\} $ to create bootstrapped mixtures:
\begin{equation}
\label{eq:empirical_mean_student_inference}
    \begin{gathered}
    \widetilde{s}^*, \widetilde{n}^* = f_{\mathcal{T}}(m^*=s^*+n^*; \w_{\mathcal{T}}) \\
    \widetilde{s}_b', \widetilde{n}_b' = f_{\mathcal{T}}(m_b'=s_b'+n_b'; \w_{\mathcal{T}})\\ 
    \widetilde{m}_b' = \widetilde{s}^* + \widetilde{n}_b', \enskip \forall b \in \{1, \dots, B\}.
    \end{gathered}
\end{equation}
In the student-training phase, we perform inference using the student network $f_{\cal S}$ on the batch of the aforementioned bootstrapped mixtures $\widetilde{m}_b'$. Because \textit{RemixIT}'s loss is computed under expectation (Equation \ref{eq:equiv_student_train}), we can rearrange the order of batches that the student network sees. Thus, we focus on the learning aspect of the student network for the batch of bootstrapped mixtures above (Equation \ref{eq:empirical_mean_student_inference}) and rewrite the last error correlation term as follows:
\begin{equation}
\label{eq:empirical_mean_student_error}
    \begin{gathered}
    \E \left[ \langle \widehat{\mathbf{R}}_{\cal S}, \widetilde{\mathbf{R}}_{\cal T} \rangle \right] \approx \E \left[
    \frac{1}{B} {\textstyle \sum}_{b=1}^B \lp \widehat{s}_b - s^* \rp^{\T} \lp \widetilde{s}^* - s^* \rp \right] \\ = \E [
    \lp \widetilde{s}^* - s^* \rp^{\T} \underbrace{\frac{1}{B} {\textstyle \sum}_{b=1}^B \lp f^{(\widehat{s})}_{\mathcal{S}}(\widetilde{s}^* + \widetilde{n}_b'; \w_{\mathcal{S}}) - s^* \rp}_{\text{\small Empirical mean student error}} ]
    \end{gathered}
\end{equation}
The premise is that if the student sees a wide variety of bootstrapped mixtures which have been generated using the same teacher's speech estimate $\widetilde{s}^*$, then the mean interference error produced by injecting noisy teacher's estimates $\widetilde{n}_b'$ would go to zero under expectation. We prove this claim under ideal conditional independence of the student error vectors and infinite bootstrapped mixtures in Theorem \ref{theorem:remixit_convergence}. In practice, the student could still minimize the errors' correlation term and still be able to learn from mixtures when the teacher performs poorly (please see Section \ref{sec:results:robust_learning} which gives an empirical analysis of our claim). 

\begin{theorem}
\label{theorem:remixit_convergence}
Assuming a differentiability of the loss functions, access to infinite bootstrapped mixtures $B \rightarrow \infty$ generated by the teacher network $f_{\T}$, and conditional independence of the student errors given the same teacher speech pseudo-target ($f^{(\widehat{s})}_{\mathcal{S}}(\widetilde{s}^* + \widetilde{n}_i'; \w_{\mathcal{S}}) - \widetilde{s}^* \bot f^{(\widehat{s})}_{\mathcal{S}}(\widetilde{s}^* + \widetilde{n}_j'; \w_{\mathcal{S}}) - \widetilde{s}^*$ with $i \neq j$), then the gradients of \textit{RemixIT}'s loss function w.r.t. the student network weights $\w_{\mathcal{S}})$ converge to the ones provided by an oracle supervised loss $ \nabla_{\w_{\mathcal{S}}}\L_{\operatorname{RemixIT}} \approx  \nabla_{\w_{\mathcal{S}}}\L_{\operatorname{Supervised}}$
\end{theorem}

\begin{proof}
Combining the definitions of the loss functions from Equations \ref{eq:supervised_loss} and \ref{eq:student_train}, their difference can be expressed as:
\begin{equation}
\label{eq:proof_remix_sup_diff}
    \mathcal{L}_{\operatorname{RemixIT}} - \mathcal{L}_{\operatorname{Supervised}} = \E \left[ || \widetilde{\mathbf{R}}_{\cal T} ||^2_2 
    - 2  \langle \widehat{\mathbf{R}}_{\cal S}, \widetilde{\mathbf{R}}_{\cal T} \rangle \right].
\end{equation}
Following the same analysis with Section \ref{sec:remix_method:remixit:error_analysis}, for each target speaker waveform $s^*$, we use the estimates of the teacher model for the target speech waveform $\widetilde{s}^*$ in an input mixture $m^*$ and for randomly sampled noise sources $\widetilde{n}_b'$ in the corresponding mixtures $m_b'$ as in Equations \ref{eq:bootstrapped_mixtures} to produce bootstrapped mixtures $\widetilde{m}_b' = \widetilde{s}^* + \widetilde{n}_b', \enskip \forall b$. Thus, the student estimates some speech waveform $\widehat{s}_b$ for each input bootstrapped mixture $\widetilde{m}_b'$ and the latter term of the error correlation can be written as follows:
\begin{equation}
\label{eq:proof_error_corr}
\begin{gathered}
\E \left[ \langle \widehat{\mathbf{R}}_{\cal S}, \widetilde{\mathbf{R}}_{\cal T} \rangle \right] =  \E \left[ \lp \widetilde{s}^* - s^* \rp^{\T} \frac{1}{B} {\textstyle \sum}_{b=1}^B \lp \widehat{s}_b - s^* \rp \right] = \\
\E \left[ \lp \widetilde{s}^* - s^* \rp^{\T} \frac{1}{B} {\textstyle \sum}_{b=1}^B \left[ ( \widehat{s}_b - \widetilde{s}^*) + (\widetilde{s}^* - s^*) \right] \right] = \\
\E \left[ || \widetilde{\mathbf{R}}_{\cal T} ||^2_2 \right] + \E \left[ \lp \widetilde{s}^* - s^* \rp^{\T} \frac{1}{B} {\textstyle \sum}_{b=1}^B \lp \widehat{s}_b - \widetilde{s}^* \rp \right].
\end{gathered}
\end{equation}
However, the error between each pseudo-target provided by the teacher student $\widetilde{s}^* = f^{(\widetilde{s})}_{\mathcal{T}}(m^*=s^*+n^*; \w_{\mathcal{T}})$ and the estimated speech signal by the student $\widehat{s}_b' = f^{(\widehat{s})}_{\mathcal{S}}(\widetilde{s}^* + \widetilde{n}_b'; \w_{\mathcal{S}})$ is bounded for any masked-based network operating on some linear bases (we use a linear encoder/decoder as specified in Section \ref{sec:exp_fram:model})  \cite{tzinis2020twostep}. Formally, assuming that an the encoded representation of the input bootstrapped mixture $m'$ is $v' = \mathbb{P} \cdot m'$, then the latent representation of a signal estimate is $\hat{v} = \widehat{\mathbf{M}} \odot (\mathbb{P} \cdot m')$. Thus, the $l_2$ error is bounded by:
\begin{equation}
\label{eq:proof_bounded_student_teacher_error}
\begin{gathered}
\|\widehat{s}_b' - \widetilde{s}^* \| = \| (\widehat{\mathbf{M}}_b' - \widetilde{\mathbf{M}}^*) \odot ( \mathbb{P} \cdot \widetilde{m}_b' ) \| \\ \leq  \underset{\widetilde{s}^*, \widehat{s}_b', \widetilde{n}_b'}{\max} \left[ \sigma_{\operatorname{max}} \{ (\widehat{\mathbf{M}}_b' - \widetilde{\mathbf{M}}^*) \odot \mathbb{P} \} \cdot \| \widetilde{s}^* + \widetilde{n}_b' \| \right] = \widehat{C},
\end{gathered}
\end{equation}
where $\sigma_{\operatorname{max}} \{ (\widehat{\mathbf{M}}_b' - \widetilde{\mathbf{M}}^*) \odot \mathbb{P} \} < \infty$ denotes the maximum singular value of the masked unrolled synthesis-basis matrix $\mathbb{P}$ and $\| \widetilde{s}^* + \widetilde{n}_b' \| < \infty$ is the energy of the bounded-norm bootstrapped mixtures. Similarly, the teacher error is also bounded by some real value $\|\widetilde{s}^* - s^*\| \leq \widetilde{C} < \infty, \enskip \forall s^*$.

Thus, by combining the above inequalities with Equations \ref{eq:proof_remix_sup_diff} and \ref{eq:proof_error_corr}, we conclude that at the limit, the difference of the loss functions converges to a value which is constant with respect to the student network's parameters $\w_{\mathcal{S}}$ as shown next:
\begin{equation}
\label{eq:proof_diff_limit}
\begin{gathered}
\lim_{B \rightarrow \infty} \left[ \mathcal{L}_{\operatorname{RemixIT}} - \mathcal{L}_{\operatorname{Supervised}} \right] = -\E \left[ || \widetilde{\mathbf{R}}_{\cal T} ||^2_2 \right]  \\ 
+ \lim_{B \rightarrow \infty} \E \left[ \lp \widetilde{s}^* - s^* \rp^{\T} \frac{1}{B} \sum_{b=1}^B \lp \widehat{s}_b - \widetilde{s}^* \rp \right] = \\
-\E \left[ || \widetilde{\mathbf{R}}_{\cal T} ||^2_2 \right] + \lim_{B \rightarrow \infty} \underset{\widetilde{s}^*}{\E} \left[ \lp \widetilde{s}^* - s^* \rp^{\T} \mu(\widetilde{s}^*) \right].
\end{gathered}
\end{equation}
where the last step comes from the application of the central limit theorem since by assumption the student estimates' errors are i.i.d. and bounded, thus, the sample mean converges in distribution to a normal distribution with mean equal to the mean student error $\mu(\widetilde{s}^*)$ given the corresponding teacher's speech estimate $\widetilde{s}^*$. All parts of the right hand-side of the above equation are constant w.r.t. the student network parameters $\w_{\mathcal{S}}$ which we try to optimize and thus by applying the gradient operator we can conclude that $ \nabla_{\w_{\mathcal{S}}}\L_{\operatorname{RemixIT}} \approx  \nabla_{\w_{\mathcal{S}}}\L_{\operatorname{Supervised}}$. One could make this theorem even more applicable to real-world settings where the student errors given different bootstrapped mixtures from the initial teacher estimate $\widetilde{s}^*$ are weakly dependent \cite{fleermann2022clt_weakly_dependent} but we defer this derivation to future work.
\end{proof}

\section{Experimental Framework}
\label{sec:exp_fram}

\subsection{Datasets}
\label{sec:exp_fram:datasets}

\noindent \textbf{DNS-Challenge (DNS)}: The DNSChallenge 2020 benchmark dataset \cite{dnschallenge_2020} consists of a large collection of clean speech recordings which are mixed with a wide variety of noisy speech samples with $64{,}649$ and $150$ pairs of clean speech and noise recordings for training and testing, respectively. DNS is used for showing the effectiveness of the proposed self-training scheme where large amounts of unsupervised training data is available and one needs to improve the performance of a model trained only on limited OOD supervised data.

\noindent \textbf{LibriFSD50K (LFSD)}: This data collection includes $45{,}602$ and $3{,}081$ mixtures for training and testing, correspondingly. The clean speech samples are drawn from the LibriSpeech \cite{librispeech} corpus and the noise recordings are taken from FSD50K \cite{fsd50k} representing a set of almost $200$ classes of background noises after excluding all the human-made sounds from the AudioSet ontology \cite{audioset}. A detailed recipe of the dataset generation process is presented in \cite{tzinis2021separate}. LFSD becomes an ideal candidate for semi-supervised/SSL teacher pre-training on OOD data given its mixture diversity.

\noindent \textbf{WHAM!}: The generation process for this dataset produces $20{,}000$ training noisy-speech pairs and $3{,}000$ test mixtures from the initial WHAM! \cite{WHAM} dataset and has been identical to the procedure followed in  \cite{tzinis2021separate} with active noise sources mixed at an average of $-1.3$dB input SNR. The set of background noises in WHAM! is limited to 10 classes of urban sounds.

\noindent \textbf{VCTK}: The VCTK dataset proposed in \cite{giri2019attention_VCTKDEMAND_testSet} includes $586$ synthetically generated noisy test mixtures, where a speech sample from the VCTK speech corpus \cite{vctk} is mixed with an isolated noise recording from the DEMAND \cite{demand_Dataset}. The VCTK and DNS test partitions are used to illustrate the effectiveness of \textit{RemixIT} under a restrictive scenario zero-shot domain adaptation with limited data to perform self-training.

\subsection{Speech enhancement model}
\label{sec:exp_fram:model}
In the supervised and \textit{RemixIT} training recipes, the student has $M=2$ output slots and always estimates the speech component and the noise source. For the models which are trained with MixIT, we increase the number of output slots to $M=3$ to estimate the additional noise component. \textit{RemixIT} is independent of the choice of the speech-enhancement model architecture as long as the latter estimates both speech and noise components of the input mixture. 

Our model's choice was based on obtaining adequate quality of speech reconstruction with low computational and memory requirements (see Table \ref{tab:final_table} for a head-to-head comparison in a supervised in-domain training setup with the previous state-of-the-art model). To this end, we used the Sudo rm -rf \cite{tzinis2020sudo} architecture with the more sparse computation blocks using shared sub-band processing via group communication \cite{luo2021ultra}. The selected network has shown to provide high-quality source estimates under speech enhancement \cite{tzinis2021separate} as well as sound separation \cite{tzinis2021compute} tasks while significantly reducing the model's size. We consider the selected architecture with a default encoder/decoder with $512$ basis $41$ filter taps and a hop-size of $20$ time-samples, a depth of $U=8$ U-ConvBlocks and the same parameter configurations as used in \cite{tzinis2021separate}. In the sequential update protocol we increase the depth of the new student networks every $20$ epochs from $8$ to $16$ and finally $32$.

\subsection{\textit{RemixIT}'s teacher update protocols configurations}
\label{sec:exp_fram:remixit_coinfigs}
\textit{RemixIT} refines the estimates of the student network based on unsupervised and semi-supervised teachers pre-trained on an OOD dataset but also has the capability of repeatedly updating the teacher network to learn from higher-quality source estimates. In our experiments, we evaluate the proposed method under various online teacher updating protocols after $k$ training epochs. Specifically, we consider the following:

\noindent\textbf{Static teacher}: The teacher is frozen throughout the training process, for all optimization steps $\w^{(k)}_{\mathcal{T}} = \w^{(0)}_{\mathcal{T}}, \enskip \forall k$.

\noindent \textbf{Sequentially updated teacher}: Every 20 epochs or equivalently $K = \nicefrac{ 20 \times | \D_m |}{B}$ optimization steps, where  $\nicefrac{ | \D_m |}{B}$ is the number of batches per-training epoch, we replace the teacher with the latest student, namely, $\mathring{\w}^{(k \operatorname{mod} K)}_{\mathcal{T}} \coloneqq \w^{(k \operatorname{mod} K)}_{\mathcal{S}}$.

\noindent \textbf{Exponentially moving average teacher}: The teacher is gradually updated after every epoch using an exponential moving average scheme $\bar{\w}^{(j+1)}_{\mathcal{T}} \coloneqq \gamma \w^{(j)}_{\mathcal{S}} + (1 - \gamma) \bar{\w}^{(j)}_{\mathcal{T}},\enskip \forall k$ with $\gamma=0.01$, where $j$ is a multiple of $\nicefrac{ | \D_m |}{B}$.
\subsection{Training and evaluation details}
\label{sec:exp_fram:train_eval_details}
For the semi-supervised and unsupervised \textit{RemixIT}'s teachers we pre-train the corresponding models following the supervised training process (Section \ref{sec:remix_method:supervised_training}) and MixIT (Section \ref{sec:remix_method:mixit}), respectively. Although \textit{RemixIT} can theoretically work with any valid signal-level loss functions (Equations \ref{eq:mixit_loss}, \ref{eq:student_train}), we choose the negative scale-invariant signal to distortion ratio (SI-SDR) \cite{sisdr} for training all models: 
\begin{equation}
\label{eq:SISDR}
    \begin{gathered}
    \mathcal{L}(\widehat{y}, y) = - \text{SI-SDR}(\widehat{y}, y) = - 20 \log_{10} ( \nicefrac{\| \alpha y\|}{\| \alpha y - \widehat{y}\|} ).
    \end{gathered}
\end{equation}
$\alpha =  \widehat{y}^\top  y /\|y\|^2$ makes the loss invariant to the scale of the estimated source $\widehat{y}$ and the target signal $y$. By setting $\alpha=1$, SI-SDR becomes equivalent with SNR. We train all models using the Adam optimizer \cite{adam} with a batch size of $B=2$ and an initial learning rate of $10^{-3}$ which is divided by $2$ every $6$ epochs. We fix those hyper-parameters after some early experimentation with the validation set of LFSD. For all experiments, during training we assume that we do not have access to the input SNR distribution and thus, we mix a clean and a noise source without altering their corresponding power ratio. However, for the in-domain supervised training setup with DNS we randomly mix clean speech and noise recordings with SNR from a uniform distribution of $[-2, 20]dB$, which has been shown to be effective for multiple sound separation setups \cite{tzinis2020twostep,zeghidour2021wavesplit,subakan2021attention}. Finally, we normalize all input mixture waveforms by subtracting their mean and dividing by their standard deviation before feeding them to each model. We train and test models which operate at a $16$kHz sampling rate.

The robustness of all speech enhancement models is measured using the SI-SDR \cite{sisdr}, the short-time objective intelligibility (STOI) \cite{stoi} and the perceptual evaluation of speech quality (PESQ) \cite{pesq}. We evaluate the model checkpoints after $100$ epochs for the pre-trained teachers and the supervised models and after $60$ epochs for all the other configurations.

\begin{table*}[htb!]
    \centering
    \begin{tabular}{ll|c|cc|cc|cc|ccc}    
    \hlinewd{1pt}
    \multicolumn{2}{l|}{\multirow{3}{*}{Training method and model details}}  & $\#$Model  & \multicolumn{6}{c|}{Available Training Data (\%)} & \multicolumn{3}{c}{Mean evaluation metrics}  \\
    & &  Params & \multicolumn{2}{c|}{Clean Speech $\D_s$} & \multicolumn{2}{c|}{Noise $\D_n$} & \multicolumn{2}{c|}{Mixture $\D_m$} & SISDR & \multirow{2}{*}{PESQ}  & \multirow{2}{*}{STOI}  \\
    & &  ($10^6$)  & DNS & LFSD & DNS & LFSD & DNS & LFSD & (dB) & &  \\
    \hlinewd{1pt}
    \multicolumn{2}{c|}{Input Noisy Mixture} & - & & & & & & & 9.2 & 1.58 & 0.915 \\
    \hlinewd{1pt}
    \multirow{3}{1.875cm}{MixIT with Sudo rm-rf ($U=8$)} & In-domain noise & $0.79$ & & & 20\% & & 80\% & & 14.4 & 2.13 & 0.933 \\
    & OOD noise & $0.79$ & & &  & 20\% & 100\% & & 14.3 & 2.02 & 0.933 \\
    & Extra OOD noise \cite{saito2021trainingSEsystemsWNoisyDatasets} & $0.79$ & & &  & 50\% & 100\% & & 14.5 & 2.03 & 0.930 \\
    \hlinewd{1pt}
    \multirow{3}{1.875cm}{Unsupervised RemixIT (ours)} & Teacher ($U=8$) & $0.79$ & & & & 20\% & & 80\% & 14.8 & 2.15 & 0.940 \\
    & Student ($U=8$) & $0.56$ & & & &  & 100\% &  & 15.5 & 2.27 & 0.947 \\
    & Student ($U:8 \rightarrow 16 \rightarrow \mathbf{32}$) & $0.73$ & & & & & 100\% & & 16.0 & 2.34 & 0.952 \\
    \hlinewd{1pt}
    \multirow{3}{1.875cm}{Semi-supervised RemixIT (ours)} & Teacher ($U=8$) & $0.56$ & & 100\% & & 100\% & & & 17.6 & 2.61 & 0.958 \\
    & Student ($U=8$) & $0.56$ & & & &  & 100\% &  & 17.6 & 2.52 & 0.956 \\
    & Student ($U:8 \rightarrow 16 \rightarrow \mathbf{32}$) & $0.73$ & &  & & & 100\% & & 18.0 & 2.60 & 0.959 \\
    \hlinewd{1pt}
    Supervised & FullSubNet$^*$ \cite{hao2021fullsubnet} & $5.6$ & 100\% & & 100\% & & & & 17.3 & 2.78 & 0.961 \\
    in-domain & Sudo rm -rf \cite{tzinis2021compute} ($U=8$) & $0.56$ & 100\% & & 100\% & & & & 18.6 & 2.69 & 0.962 \\
    training & Sudo rm -rf \cite{tzinis2021compute} ($U=32$) & $0.73$ & 100\% & & 100\% & & & & \textbf{19.7} & \textbf{2.95} & \textbf{0.971} \\
    
    \bottomrule
    \end{tabular}
    \caption{Evaluation results for the speech enhancement task on the DNS test set using the proposed \textit{RemixIT} method, MixIT approaches \cite{mixit,saito2021trainingSEsystemsWNoisyDatasets} and supervised in-domain training with the Sudo rm -rf model \cite{tzinis2021compute} as well as the previous state-of-the-art \textit{FullSubNet} \cite{hao2021fullsubnet} supervised model ($^*$as it was presented in the paper). All teacher and student networks follow the same Sudo rm -rf model \cite{tzinis2021compute} architecture with the specified number of U-ConvBlocks ($U=8$ or $U=32$). $U:8 \rightarrow 16 \rightarrow \mathbf{32}$ denotes that we double the depth of the student network every $20$ epochs and sequentially update the teacher with the latest available student, the reported number refers to the performance of the student with $U = 32$.}
    \label{tab:final_table}
    \vspace{-10pt}
\end{table*}

\section{Results and Discussion}
\label{sec:results}
\subsection{The need for continual refinement of teacher's estimates}
\label{sec:results:continuous_refinement}
\begin{figure}[t!]
    \centering
      \includegraphics[width=\linewidth]{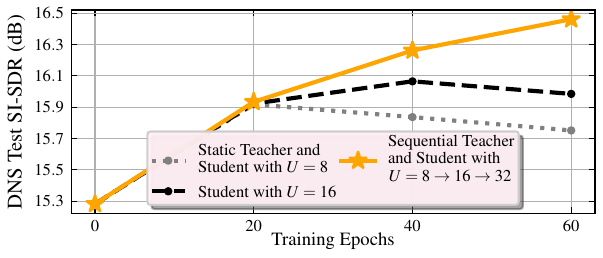}
      \caption{SI-SDR (dB) performance on DNS test as training progresses using different teacher update protocols. The initial teacher network is shared across the various protocols (a Sudo rm -rf model with $U=8$ Conv-blocks) and was pre-trained in a supervised way on the WHAM! dataset. The orange solid line denotes the performance of the student model with increasing depth every $20$ training epochs $U:8\rightarrow 16 \rightarrow 32$ where we initialize a new student model and replace the teacher model with the latest available student. The sequential protocol shows significant gains over the static teacher protocols where the student network has a static architecture throughout training and the initial teacher is not updated ($U=8$ with gray and $U=16$ with black dashed lines).} \label{fig:static_vs_sequential}
      \vspace{-10pt}
\end{figure}
In Figure \ref{fig:static_vs_sequential}, we show the speech enhancement performance of the student models produced by a static or a sequentially updated teacher every $20$ epochs. Unsurprisingly, all protocols behave similarly until the $20$th epoch since they use the same initial teacher. In contrast to the frozen teacher protocols, after the $20$th epoch, the old teacher is replaced with the newly trained student with $U$=8 and a new student, with twice as much depth (8 $\rightarrow$ 16) is initialized. Surprisingly, the sequentially updated teacher protocol keeps teaching a better student separation model, even after the $40$th epoch, compared to both models produced by the static teachers which saturate for the same number of training steps. Comparing between the students with $U$=8 and $U$=16 produced by static teacher models, it is evident that the more expressive student performs better but not on par with the same depth student produced by the sequentially updated teacher protocol. Specifically, both orange-solid and black-dashed lines at the $40$-th epoch represent the performance of a student model with the same depth ($U=16$) but the sequential update protocol clearly outperforms the frozen-teacher protocol. Thus, the combination of the bootstrapped remixing and the continual refinement of the teacher's estimates is key for the significant improvement that \textit{RemixIT} yields. As a result, we have chosen the sequentially updated teacher protocol as the default strategy for \textit{RemixIT}, except for the zero-shot adaptation where we use the exponential average teacher updating scheme because the number of available training mixtures could make the student prone to overfitting if trained from scratch. 

\subsection{Self-supervised and semi-supervised speech enhancement}
\label{sec:results:self_sup}

Table \ref{tab:final_table} summarizes the mean speech enhancement performance of \textit{RemixIT} against in-domain and cross-domain supervised and SSL baselines with the same architecture on the DNS test set. Notice that in both semi-supervised and unsupervised cases, the learned \textit{RemixIT}'s student does not assume access to in-domain clean speech nor to noise samples like the previous state-of-the-art SSL speech enhancement algorithms. For instance, SSL \textit{RemixIT}'s teacher pre-training is performed with OOD MixIT by using $80\%$ of the LFSD noisy recordings $\D_m'$ and rest $20\%$ to simulate the isolated noise recordings $\D_n'$, whereas the student is trained solely on the vast amount of training mixtures in the DNS dataset. 

Despite the fact that \textit{RemixIT} makes no assumptions about the in-domain distribution of mixtures nor it assumes access to in-domain ground truth source waveforms, it significantly outperforms all the previous state-of-the-art MixIT-like approaches. The unsupervised student learned using the proposed method yields an improvement over the second-best unsupervised method of more than ($14.5$dB $\rightarrow 16.0$dB in terms of SI-SDR and $0.02$ in terms of STOI) compared against in-domain MixIT and the recently proposed extra noise augmentation where an extra noise source is injected \cite{saito2021trainingSEsystemsWNoisyDatasets}. In the semi-supervised domain adaptation setup, we show that \textit{RemixIT}'s student still provides noticeable improvement over its initial teacher pre-trained in a supervised way assuming access to a smaller but diverse dataset like LFSD. Although we have used the same separation model architecture across our experiments, our method is independent of the model's choice and could be used with models that produce higher quality estimates. However, the bottom three rows in Table \ref{tab:final_table} show that the model used in this study achieves state-of-the-art speech enhancement results when trained with in-domain ground-truth sources.

\begin{figure*}[ht]
    \centering
  \begin{subfigure}[h]{\linewidth}
    \centering
      \includegraphics[width=0.9\linewidth]{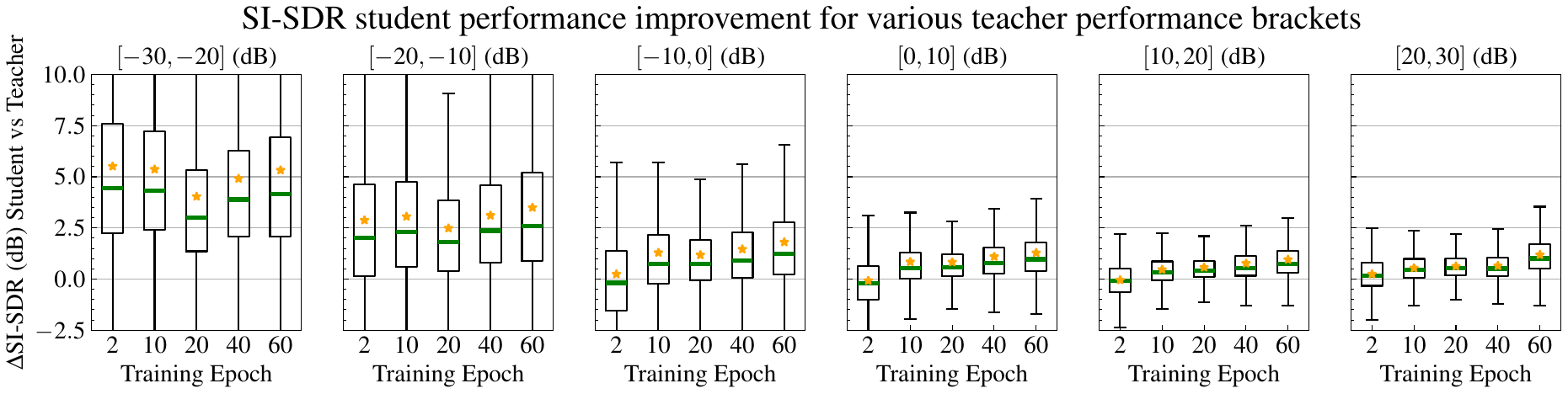}
      \caption{Semi-supervised \textit{RemixIT} with initial teacher pre-trained on WHAM! in a supervised way.}
      \label{fig:semi_sup_learning_progression} 
     \end{subfigure}
  \begin{subfigure}[h]{\linewidth}
  \centering
\includegraphics[width=0.9\linewidth]{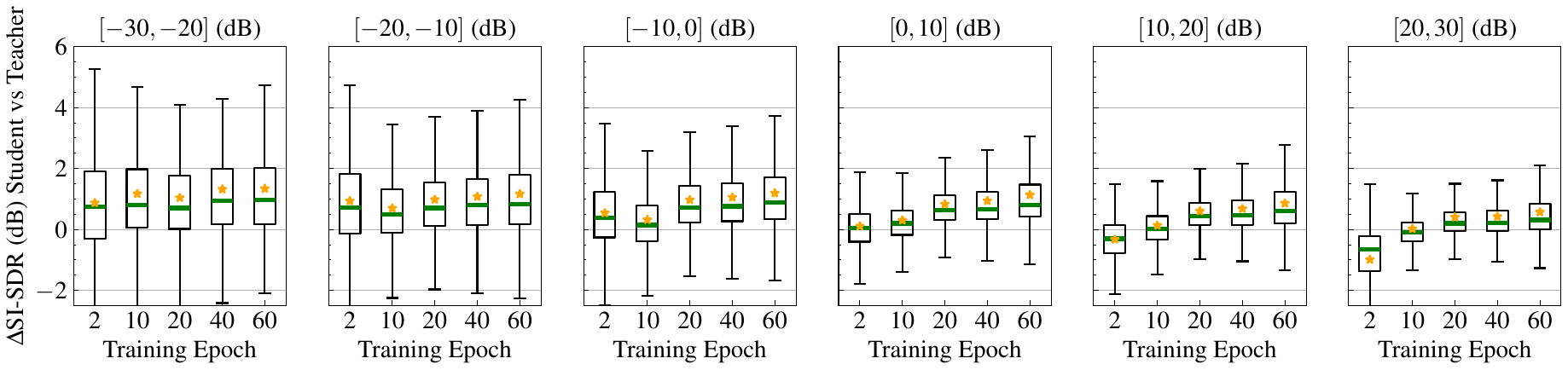}
      \caption{Unsupervised \textit{RemixIT} with initial teacher pre-trained on LFSD using MixIT.}
      \label{fig:unsup_learning_progression} 
     \end{subfigure} \\
     \caption{SI-SDR (dB) performance improvement on the training portion of the DNS dataset that a \textit{RemixIT}'s student with a sequentially updated teacher every 20 epochs yields as the training progresses over the initial teacher's estimates. We show that similar learning patterns emerge for different initial teachers pre-trained in a semi-supervised way (top) and an unsupervised way (bottom). The median and the mean $\Delta$SI-SDR are denoted with a solid green line and an orange star, respectively. }
    \label{fig:learning_progression}
    \vspace{-10pt}
\end{figure*} 

\begin{figure}[ht]
    \centering
      \includegraphics[width=0.9\linewidth]{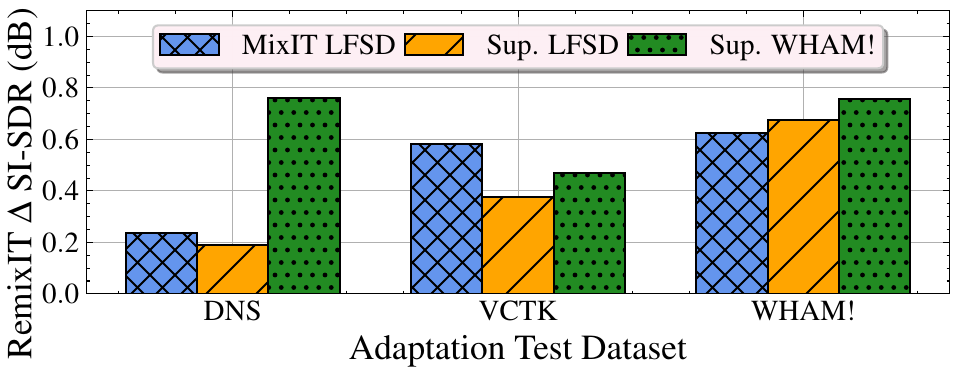}
      \caption{SI-SDR performance improvement that \textit{RemixIT}'s student yields over its initial OOD pre-trained teacher model for various low-resource adaptation datasets (e.g. DNS, LFSD and WHAM!, from left to right). Both teacher and student models have the exact same Sudo rm -rf architecture ($U=8$ ConvBlocks) and we use the running mean teacher update protocol. \textit{RemixIT} shows significant improvements against all teacher models used in this study, namely, MixIT pre-training on LFSD (blue/leftmost) and supervised training on LFSD (yellow/middle) and as well as on WHAM! (green/rightmost).}
      \label{fig:zeroshot}
      \vspace{-10pt}
\end{figure}
\subsection{Zero-shot domain adaptation}
\label{sec:results:zero_shot}
In low-resource training scenarios, the training mixtures in-hand might not be sufficient to train a model from scratch, thus, we show how \textit{RemixIT} can be used as a zero-shot unsupervised domain adaptation algorithm. We perform teacher pre-training on larger OOD datasets and fine-tune a student model using limited in-domain mixtures. At the start of the adaptation process, the student is initialized using the pre-trained teacher's weights $\w_{\mathcal{S}}^{(0)} := \w_{\mathcal{T}}^{(0)}$ and we perform \textit{RemixIT} while periodically updating the teacher using the moving average protocol (see Section \ref{sec:exp_fram:remixit_coinfigs}). The cross-dataset adaptation results are illustrated in Figure \ref{fig:zeroshot}. The proposed method delivers consistent improvements across datasets and pre-training techniques, up to $0.8$dB in terms of SI-SDR over the non-calibrated models. Unsurprisingly, one can notice that the level of improvement is directly impacted by the amount of available noisy mixtures. We postulate that this is the main reason that our method obtains larger (smaller) gains for the adaptation on WHAM! (DNS) test partition which has $3{,}000$ (only $150$) mixtures, respectively. However, \textit{RemixIT} performs adequately even in cases where there is a large distribution shift between the training and the adaptation-test sets (e.g. WHAM! contains only 10 classes of urban background noises while the DNS dataset is very diverse). Specifically, the significant improvement after adapting a supervised pre-trained model on WHAM! to the $150$ mixtures of the very diverse DNS set, indicates the effectiveness of \textit{RemixIT} under really challenging zero-shot learning conditions.
\begin{figure*}[ht]
    \centering
  \begin{subfigure}[h]{\linewidth}
  \centering
  \includegraphics[width=0.9\linewidth]{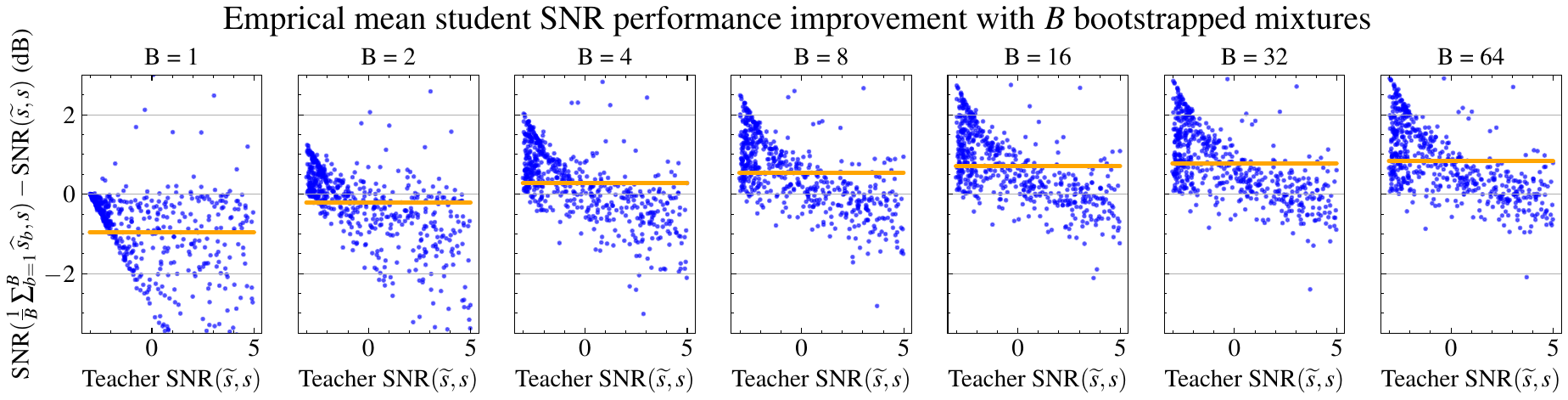}
      \caption{Semi-supervised \textit{RemixIT} with initial teacher pre-trained on WHAM! in a supervised way.}
      \label{fig:semi_sup_robust_learning} 
     \end{subfigure}
  \begin{subfigure}[h]{\linewidth}
  \centering
\includegraphics[width=0.9\linewidth]{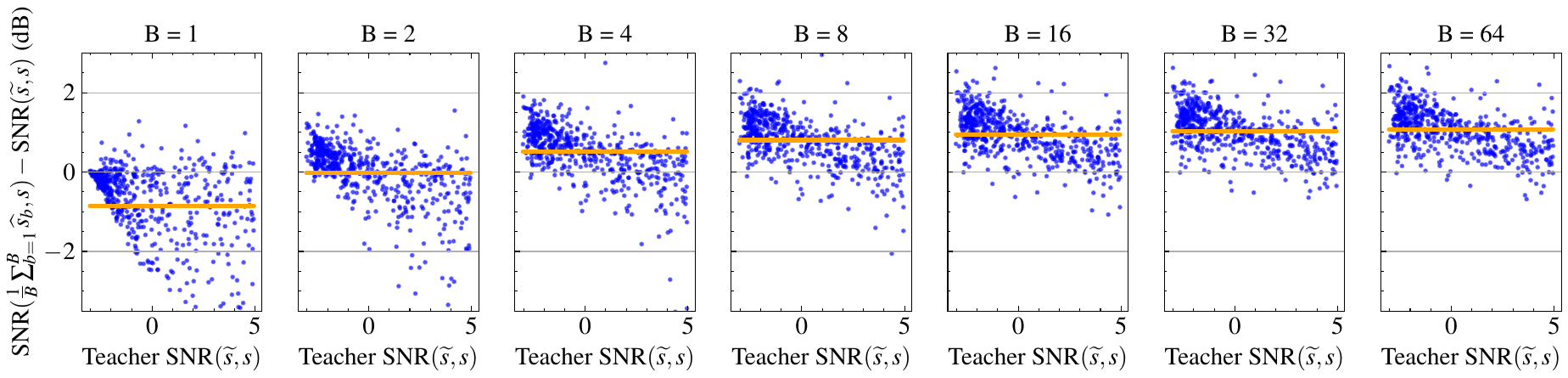}
      \caption{Unsupervised \textit{RemixIT} with initial teacher pre-trained on LFSD using MixIT.}
      \label{fig:unsup_robust_learning} 
     \end{subfigure} \\
     \caption{Distribution of SNR improvement (dB) on the DNS training set that the empirical mean \textit{RemixIT}'s student after 10 training epochs ( Equation \ref{eq:empirical_mean_student_error}) yields over its initial teacher in regions where the latter performs poorly. The solid orange line denotes the mean SNR improvement for each number of bootstrapped mixtures which are considered under expectation $\frac{1}{B} \sum_{b=1}^B f^{(\widehat{s})}_{\mathcal{S}}(\widetilde{s} + \widetilde{n}_b'; \w_{\mathcal{S}})$. We show that as a fixed student network sees more input bootstrapped mixtures, the mean student performance becomes better on average than its teacher even early in training and in regions where the teacher performs poorly. 
    }
    \label{fig:robust_learning}
    \vspace{-10pt}
\end{figure*} 
\subsection{Student learning progression}
\label{sec:results:student_learning_progression}
We analyze how a student speech enhancement model trained with \textit{RemixIT} on the DNS train set refines its estimates as the training progresses and how it compares against its initial teacher. In Figure \ref{fig:learning_progression}, we showcase the improvement obtained in terms of SI-SDR for various teachers and their performance brackets under a sequentially updated teacher every 20 epochs using the parameters from the student network. Note that the student is gradually learning to perform better than the initial teacher network in the regions where the latter performs better (rightmost plots row-wise) even if producing improvement over really good estimates (e.g. higher than $15$dB) becomes harder. Thus, it becomes evident that the continual self-training scheme of \textit{RemixIT} where the teacher network is updated using the latest student's weights is key to a larger performance boost. The result holds for both OOD supervised and MixIT teachers and is on par-with our theoretical analysis in Section \ref{sec:remix_method:remixit:error_analysis} where we show how a better teacher helps the error correlation term of \textit{RemixIT}'s loss function to diminish and resemble supervised training. In contrast, for the low performing brackets ($[-30, -10]$dB in terms of teacher SI-SDR (dB)), the student does not learn how to further increase its performance, even if it regresses over the estimated waveforms of updated teachers. The emergence of this learning pattern necessitates the discovery of more robust self-training algorithms which can recover from cases where the teacher network provides extremely noisy estimates.

\subsection{Robust learning with very noisy teacher's estimates}
\label{sec:results:robust_learning}
We investigate the robustness of \textit{RemixIT} in cases where the teacher model outputs a low quality speech estimate $\widetilde{s}$. Building upon the analysis performed in Section \ref{sec:remix_method:remixit:error_analysis}, we reiterate on how important is for the student to be trained on multiple bootstrapped mixtures $\widetilde{m}_b' = \widetilde{s} + \widetilde{n}_b', \enskip \forall b \in \{1, \dots, B\}$ produced using the same teacher's speech estimate $\widetilde{s}$ and independent teacher's noise estimates $\widetilde{n}_b'$ (see Equations \ref{eq:empirical_mean_student_inference}, \ref{eq:empirical_mean_student_error}). The distribution of the SNR performance improvement that the empirical mean student yields over the initial teacher after 10 training epochs is displayed in Figure \ref{fig:robust_learning} while sweeping the number of input bootstrapped mixtures. For both cases of supervised and MixIT teachers we see that the mean SNR improvement is around $2$ dB when increasing the number of bootstrapped mixtures $B$ from $1$ to $64$. Notably, this result holds for really bad teacher estimates, namely, less than $5$ dB and is obtained by simply performing inference over more augmentations of $\widetilde{s}$ without refining the student parameters. Assuming that all speech estimates and the ground-truth signals have unit-norm $\|s\|=\|\widetilde{s}\|=\|\widehat{s}_b\|=1$, the maximization of SNR becomes equivalent to minimizing the $l2$ norm $\operatorname{SNR}(\widehat{y}, y) \propto - \| \widehat{y} - y\|$. As a result, the mean SNR improvement of the empirical mean student leads the term $\E [ \langle \widehat{\mathbf{R}}_{\cal S}, \widetilde{\mathbf{R}}_{\cal T} \rangle ]$ closer to zero (Equation \ref{eq:empirical_mean_student_error}) and consequently, the student to learn in a more robust way, even in cases where the teacher's error term $\|\widetilde{\mathbf{R}}_{\cal T}\|$ is far from zero.

\definecolor{Gray}{gray}{0.9}
\begin{table}[htb!]
    \setlength{\tabcolsep}{7.5pt} 
    \renewcommand{\arraystretch}{1.1} 
    \centering
    \begin{tabular}{l|c|cc|ccc}   
    \hlinewd{1pt}
    \multirow{3}{*}{Method} & \multirow{3}{*}{$U$} & \multicolumn{2}{c|}{Training Data} & \multicolumn{3}{c}{Mean test-set} \\
    &   & Noise  & Mixtures & \multicolumn{3}{c}{$\operatorname{SI-SDRi}$ (dB)} \\
    & &    $\D_n$ &  $\D_m$ & D & L & W! \\
    \hlinewd{1pt}
    MixIT with & 8 & D & D &  5.2 &  6.3 & 6.6 \\
    in-domain    
    & 8$^L$ &  L & L &  5.6 &  8.5 &  6.2 \\ 
    noise & 8  & W! & W! &  4.5 & 2.3 &  \textbf{9.8} \\
    \hline
     & \multirow{4}{*}{8}  & L & \multirow{2}{*}{D} & 5.1 &  3.6 & 5.3 \\ 
     MixIT with&   & W! & & 1.2 &  -0.2 &  1.7 \\
    \hhline{~~-----}
     OOD noise &   & D & \multirow{2}{*}{W!} & -0.8 &  -6.3 & 1.8 \\
     &  & L & & 1.7 & -1.7 & 1.2 \\
    \hline
    MixIT  & \multirow{4}{*}{8} & L+L & \multirow{2}{*}{D} & 5.3 &  1.3 &  4.7 \\
    with extra &  & W!+W! & & 5.3 & 5.0 & 2.9 \\
    \hhline{~~-----}
    OOD injected  &  & D+D & \multirow{2}{*}{W!} & -3.5 & -9.8 & 3.0 \\
    noise \cite{saito2021trainingSEsystemsWNoisyDatasets} 
    &  &  L+L &  & -1.6 & -10.2 & 1.6 \\
    \hlinewd{1pt}
     & 8   & \multirow{2}{*}{L} & \multirow{2}{*}{L$\rightarrow$\textbf{D}} & 6.3 &  9.3 & 7.5 \\
    \textit{RemixIT} & 32   &  &  &\cellcolor{Gray}\textbf{6.8}&\cellcolor{Gray}\textbf{9.9}&\cellcolor{Gray}8.2\\
    \hhline{~------}
    (ours) & 8   & \multirow{2}{*}{L} & \multirow{2}{*}{L$\rightarrow$\textbf{W!}} & 5.3 & 7.5 & 6.8  \\
     & 32   & &  & 4.9 &  7.2 & 6.9 \\
    \bottomrule
    \end{tabular}
    \caption{Self-supervised training mean SI-SDR improvement (dB) over the input mixture performance for MixIT baselines and \textit{RemixIT}. The initial MixIT teacher uses a sudo rm -rf \cite{tzinis2021compute} architecture with $U=8$ blocks on the LFSD (L) dataset and is denoted with $^L$. The evaluated \textit{RemixIT}'s student models follow a sequentially updated teacher protocol where they grow in depth as: $U:8\rightarrow16\rightarrow32$ and are only trained using the corresponding \textbf{bolded} mixture dataset. Gray background colored cells denote the best performing model which did not have access to clean in-domain training data for the corresponding dataset (note that MixIT assumes access to clean in-domain noise recordings). The mean noisy input mixture SI-SDR performance is $9.2$, $6.3$ and $-1.3$ dB for DNS, LFSD, and WHAM! datasets, respectively.
    }
    \label{tab:SSL_generalization}
    \vspace{-10pt}
\end{table}

\subsection{Cross-domain generalization}
\label{sec:results:cross_domain_generalization}
A comparison for cross-domain generalization in self-supervised and semi-supervised domain adaptation speech enhancement tasks is displayed in Tables \ref{tab:SSL_generalization} and \ref{tab:semisup_adaptation}, respectively. 

In Table \ref{tab:SSL_generalization}, we notice that MixIT and its variants fail to generalize in cases where the noise distribution $\D_n$ does not closely resemble the true in-domain distribution $\D_n^*$. Notably, \textit{RemixIT} outperforms all MixIT methods without having access to in-domain datasets. For instance, in the case where one only has access to mixtures from the WHAM! (W!) dataset and noise sources from LFSD (L), the best noise augmented MixIT model obtains only $1.6$ dB of SI-SDR improvement on the adaptation WHAM! dataset. In stark contrast, \textit{RemixIT} with a pre-trained MixIT teacher on LFSD yields an improvement of 5.3dB ($1.6 \rightarrow 6.9$) over the best cross-dataset trained MixIT model and 0.7 dB ($6.2 \rightarrow 6.9$) over the teacher model. 

In very harsh mismatched cases, such as when using noise samples from LFSD and mixture samples from LFSD and WHAM!, \textit{RemixIT} shows strong results for all datasets ($4.9$ dB for DNS, $7.2$ dB for LFSD and $6.9$ dB for WHAM!) while even the best MixIT configuration fails to produce significant improvements over the input mixture ($1.7$ dB for DNS, $-1.7$ dB for LFSD and $1.6$ dB for WHAM!). Moreover, \textit{RemixIT} can also improve the performance of a teacher model in the source dataset test-set by leveraging other target mixture datasets. Notice that \textit{RemixIT} yields an improvement of 1.4dB ($8.5 \rightarrow 9.9$) on the LFSD test-set over the pre-trained teacher model on LFSD by using self-training over the diverse unsupervised DNS mixture dataset. Surprisingly, \textit{RemixIT} also outperforms its teacher by a large margin ($6.2 \rightarrow 8.2$ dB) on the WHAM! test set even though it has not seen any data from this dataset which shows how \textit{RemixIT} can provide a seamless solution to generalizing denoising models to unseen data.

Although \textit{RemixIT} shows a small performance degradation in the adaptation $L \rightarrow \textbf{W!}$ set compared to the adaptation with cleaner datasets, such as: $L \rightarrow \textbf{D}$ ($7.5 \rightarrow 6.8$ for the shallow student and $8.2 \rightarrow 6.9$ for the $U=32$ student on W!), notice that the same MixIT configuration suffers a major hit in denoising performance ($5.3 \rightarrow 1.2$ dB on W!) which makes it almost similar to a no-processing model. In essence, the input SNR of WHAM! ($-1.3$dB) prevents self-supervised algorithms from learning effectively and training on OOD but higher input-SNR datasets (e.g. DNS) leads to better results.

In Table \ref{tab:semisup_adaptation}, we show that \textit{RemixIT} aptly performs semi-supervised domain adaptation even for severely mismatched cases such as transferring knowledge from DNS to the much less diverse and lower input-SNR WHAM!. \textit{RemixIT} yields the best performing model without clean in-domain source signals on the DNS test set ($7.3$ dB) when only starting from the OOD semi-supervised teacher on WHAM! with a much inferior performance of $6.1$ dB.

\begin{table}[htb!]
    \setlength{\tabcolsep}{5.5pt} 
    \renewcommand{\arraystretch}{1.1} 
    \centering
    \begin{tabular}{l|c|ccc|ccc}   
    \hlinewd{1pt}
    \multirow{3}{*}{Method} & \multirow{3}{*}{$U$} & \multicolumn{3}{c|}{Training Data} & \multicolumn{3}{c}{Mean test-set} \\
    & & Speech  & Noise  & Mixtures & \multicolumn{3}{c}{$\operatorname{SI-SDRi}$ (dB)} \\
    & &  $\D_s$ &  $\D_n$ &  $\D_m$ & D & L & W! \\
    \hlinewd{1pt}
    & 8$^D$ & \multirow{2}{*}{D} & \multirow{2}{*}{D} & & 9.4 &  10.3 &  8.4 \\
    Super-& 32 &  &  & &  \textbf{10.5} &\cellcolor{Gray}\textbf{12.2}&\cellcolor{Gray}10.2\\
    \hhline{~-------}
    vised& 8$^{W!}$ & \multirow{2}{*}{W!} & \multirow{2}{*}{W!} & &  6.1 & 4.8 & 11.3 \\
    & 32 &  &  & &  7.1 &  6.2 &  \textbf{12.8} \\
    \hlinewd{1pt}
    & 8 & \multirow{2}{*}{D} & \multirow{2}{*}{D} & \multirow{2}{*}{\textbf{W!}} & 6.9 & 5.6 & 9.0 \\
    \textit{RemixIT} & 32 &  &  &  &  6.9 & 5.4 & 9.8 \\
    \hhline{~-------}
    (ours) & 8 & \multirow{2}{*}{W!} & \multirow{2}{*}{W!} & \multirow{2}{*}{\textbf{D}} & 6.7 & 5.7 &  11.3 \\
     & 32 & & &  &\cellcolor{Gray}7.3& 6.4 &  10.9 \\
    \bottomrule
    \end{tabular}
    \caption{\textit{RemixIT} mean SI-SDR improvement (dB) over the input mixture for semi-supervised domain adaptation. The initial OOD supervised teachers with a sudo rm -rf \cite{tzinis2021compute} with $U=8$ blocks on the DNS (D) and the WHAM! (W!) datasets are denoted with $^D$ and $^{W!}$, respectively. The evaluated \textit{RemixIT}'s student models follow a sequentially updated teacher protocol where they grow in depth as: $U:8\rightarrow16\rightarrow32$ and are only trained using the corresponding \textbf{bolded} mixture dataset. Gray background colored cells denote the best performing model which did not have access to clean in-domain training data for the corresponding dataset. The mean noisy input mixture SI-SDR performance is $9.2$, $6.3$ and $-1.3$ dB for DNS (D), LFSD (L), and WHAM! (W!) datasets, respectively.
    }
    \label{tab:semisup_adaptation}
    \vspace{-10pt}
\end{table}

\subsection{RemixIT with in-domain noise recordings}
\label{sec:results:extra_clean_noise}

Finally, we also propose an extension to our proposed self-training method to adopt readily available isolated in-domain noise recordings $\n \sim \D_n$ which can further enhance \textit{RemixIT}'s performance. To do so, we alter the bootstrapped remixing process presented in Equation \ref{eq:bootstrapped_mixtures} using a portion of the in-domain noise recordings $\n \sim \D_n$ instead of the teacher's noise estimates $\ten \sim f^{\ten}_{\mathcal{T}}(\mix; \w^{(k)}_{\mathcal{T}})$ as shown below:
\begin{equation}
\label{eq:extra_noise_rec}
    \begin{gathered}
    \tmix_b = \tes_b + \zeta \n_b + (1 - \zeta) \ten^{(\textbf{P})}_b, \enskip \zeta \sim \operatorname{Bernoulli}(p_n),
    \end{gathered}
\end{equation}
where $b$ indicates the batch-index and $p_n$ is the Bernoulli parameter of sampling an in-domain noise recording instead of a teacher's estimate for the corresponding batch-index.

\begin{figure*}[ht]
    \centering
  \begin{subfigure}[h]{0.32\linewidth}
    \centering
      \includegraphics[width=\linewidth]{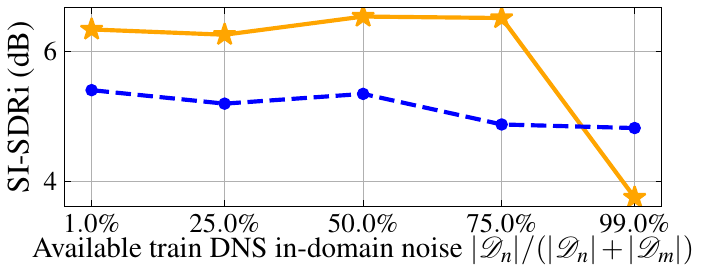}
      \caption{Performance on the DNS test set.}
      \label{fig:extra_noise_remixit:dns} 
     \end{subfigure}
  \begin{subfigure}[h]{0.32\linewidth}
    \centering
      \includegraphics[width=\linewidth]{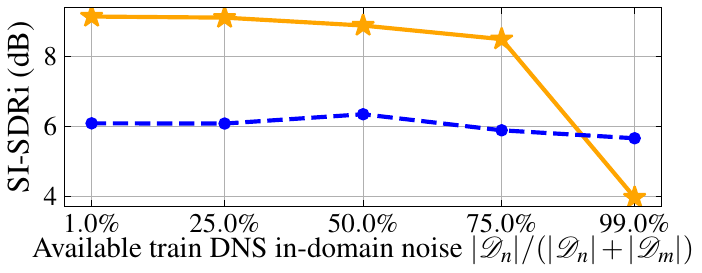}
      \caption{Performance on the LFSD test set.}
      \label{fig:extra_noise_remixit:lfsd} 
     \end{subfigure}
     \begin{subfigure}[h]{0.32\linewidth}
    \centering
      \includegraphics[width=\linewidth]{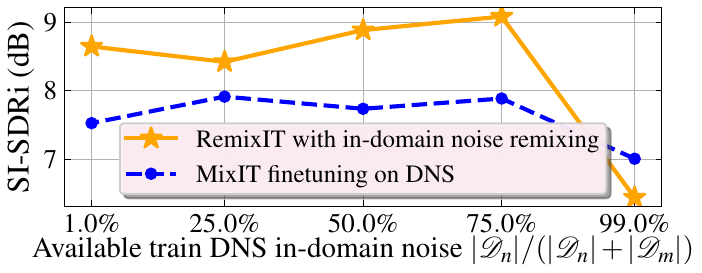}
      \caption{Performance on the WHAM! test set.}
      \label{fig:extra_noise_remixit:wham} 
     \end{subfigure}\\
     \caption{SI-SDR (dB) performance improvement of a sudo rm -rf ($U=8$) model fine-tuned using MixIT (blue-dashed line) and trained using \textit{RemixIT} with in-domain noise recordings recordings remixing (solid orange line) on different test sets with different DNS training set splits. For each plot the x-axis denote the split between the DNS-training partition between in-domain noise recordings $\D_n$ and mixture available data $\D_m$. Both self-supervised speech enhancement methods start using the same pre-trained MixIT sudo rm -rf ($U=8$) model with $20\%$ in-domain isolated noise data and $80\%$ mixture recordings from the LFSD dataset. For each method we evaluate the corresponding checkpoints that lead to the best performance on the LFSD test set after $20$ full training epochs. }
    \label{fig:extra_noise_remixit}
    \vspace{-10pt}
\end{figure*}

In Figure \ref{fig:extra_noise_remixit} we show how our method performs against a stronger fine-tuned MixIT baseline using the same Sudo rm -rf architecture with $U=8$ U-ConvBlocks on various splits of the DNS training data. The pre-trained model on LFSD data is used as an initialization checkpoint for MixIT fine-tuning and as the teacher network for performing \textit{RemixIT} with bootstrapped mixtures from teacher's estimates and in-domain noise recordings. We set the probability of synthesizing a bootstrapped mixture with an isolated in-domain noise recording instead of a teacher's noise estimate equal to the ratio of the in-domain noise recordings compared to the mixture data $p_n = \nicefrac{|\D_n|}{(|\D_n| + |\D_m|)}$. We notice that \textit{RemixIT} performs consistently better than the fine-tuned MixIT for the same ratio of in-domain noise recordings except of the rightmost point where the bootstrapped mixtures contain less diverse mixtures leading the student model to overfit to only a small amount of human utterances. Notably, \textit{RemixIT} trains a full student model from scratch compared to the fine-tuned MixIT which has more trainable parameters (0.97 millions vs 0.56) and also enjoys the warm-start from a LFSD MixIT checkpoint. It is also evident that our proposed \textit{RemixIT} extension becomes better with more supervised data for the generalization datasets (see Figure \ref{fig:extra_noise_remixit:dns} for DNS and Figure \ref{fig:extra_noise_remixit:wham} for WHAM!). This is also reflected on a small ablation study that we performed to set the in-domain noise sampling prior parameter $p_n$ in which we kept the amount of in-domain noise recordings and mixture data equal $|\D_n|=|\D_m|$ and gradually increased the Bernoulli parameter $p_n: 0.01 \rightarrow 0.5$. As a result, we noticed a performance increase in terms of SI-SDRi of $6.1 \rightarrow 6.4$ (dB) for the DNS test-set and $8.6 \rightarrow 9.0$ (dB) for the WHAM! dataset which enhances our claim that cleaner noise estimates can lead to stronger gains through synthesizing bootstrapped mixtures with less interference. 

\section{Conclusion}
\label{sec:conclusion}
We have presented a self-training scheme for speech enhancement models which is based on a lifelong bi-directional parameter update between a teacher and a student network. The proposed framework aptly transfers the knowledge of a pre-trained model on out-of-domain data using bootstrapped remixing and through the continual refinement of the teacher's outputs. We have experimentally shown that our method significantly outperforms all previous state-of-the-art self-supervised methods while being more general and without the dependence on in-domain data. Moreover, our results illustrated that \textit{RemixIT} can also perform semi-supervised and zero-shot domain adaptation setups with limited in-domain mixtures. Furthermore, our theoretical analysis is backed by empirical results and instrumental to the understanding of the teacher-student learning dynamics, especially in where our method can still learn with extremely noisy pseudo-target signals. In the future, we aim to strengthen the robustness of our algorithm by estimating a confidence-based proxy for the quality of the pseudo-targets \cite{fixmatch} as well as widen the applicability of our method by applying it to different domains.

\bibliographystyle{IEEEbib}
\bibliography{refs}
\end{document}